\documentclass[twocolumn]{jpsj2}
\def\A{\rm A}
\def\B{\rm B}
\def\C{\rm C}

\def\JA{J_{\A}}
\def\JB{J_{\B}}
\def\JC{J_{\C}}
\def\HA{\H_{\A}}
\def\HB{\H_{\B}}
\def\HC{\H_{\C}}
\def\v#1{\mbox{\boldmath$#1$}}

\def\H{{\cal H}}

\def\mpynn{$m$-MPYNN$\cdot$BF$_4$}
\def\simleq{\mbox{\raisebox{-1.0ex}{$\stackrel{<}{\sim}$}}}

\def\runtitle{$S=1$ Anisotropic Distorted Kagom\'e Heisenberg Antiferromagnets}

\def\runauthor{Kazuo {\sc Hida}}

\title
{
Ground State Phase Diagram of the Distorted $S=1$  Kagom\'e Heisenberg Antiferromagnets with Single-site Anisotropy
}

\author
{Kazuo {\sc Hida}
\footnote{e-mail: hida@phy.saitama-u.ac.jp}}

\inst
{
Department of Physics, Faculty of Science,\\ Saitama University, Saitama, Saitama 338-8570
}

\recdate
{
\today
}

\abst
{The ground state phase transitions in the distorted $S=1$ kagom\'e Heisenberg antiferromagnet (KHAF) with single-site anisotropy $D$ are studied by the numerical exact diagonalization method. For strong easy plane anisotropy, the hexgonal singlet solid (HSS) ground state of the uniform KHAF is destroyed and large-$D$ state is realized. The quantum phase transition between these two states is analogous to the Haldane-large-$D$ transition in the $S=1$ antiferromagnetic Heisenberg chain. The combined effect of the $\sqrt{3}\times\sqrt{3}$ lattice distortion and single site anisotropy is also investigated. The ground state  phase diagram expected from the numerical results is presented. The presence of this transition is  consistent with the HSS picture of the ground state of the uniform $S=1$ KHAF and supports its validity.}
\kword
{
kagom\'e Heisenberg antiferromagnet, single-site anisotropy, lattice distortion, hexagonal singlet solid state, large-$D$ state
}
\begin{document}

\maketitle
\section{Introduction}

The kagom\'e Heisenberg antiferromagnet (KHAF) has been extensively studied theoretically and experimentally because of the interest in the interplay of the strong thermal and quantum fluctuation and  highly frustrated nature of the lattice structure\cite{rev,ze1,ey1,nm1,wal1,mila1,sin1,kh1,kh2,kh3}. So far, most of the attempts have been focused on the ground state and low lying excitations of the uniform KHAF.

The present author proposed the hexagonal singlet solid (HSS) picture for the ground state of $S=1$ KHAF\cite{kh1} which is analogous to the valence bond solid (VBS) picture of the ground state of the $S=1$ antiferromagnetic Heisenberg chains (AFHC)\cite{aklt}. In both HSS and VBS pictures, the $S=1$ spins are decomposed into  symmetrized pairs of two $S=1/2$ spins. In  the HSS state, six $S=1/2$ spins are assigned for each hexagon and they form a 6-spin singlet state around each hexagon. This is in contrast to the VBS state in which these $S=1/2$ spins form 2-spin singlet state on each bond. The HSS state is explicitly constructed in ref. \citen{kh1} and shown to give a good variational energy. Similar physical picture has been also proposed for the $S=1$ pyrochlore system\cite{yama1}.

As a real material, Wada and coworkers\cite{wada1,awaga1,wata1}  have investigated the magnetic behavior of \mpynn  which can be regarded as the $S=1$ KHAF. Therefore, if the HSS picture of the ground state of $S=1$ KHAF is verified, this material is the first realistic example of the VBS-like state in 2-dimensional $S=1$ magnetic systems. In this context, it is quite important to check the validity of the HSS picture from various points of view. Especially, it is important to verify which kind of perturbations destroy the ground state of the unifom $S=1$ KHAF. In the previous work, we investigated the effect of the $\sqrt{3}\times\sqrt{3}$ lattice distortion\cite{kh3}. It is shown such distortion destroys the ground state of the uniform $S=1$ KHAF as the VBS state is destroyed by dimerization. The ground state in the strongly distorted regime is the lagre-HSS state which has the large scale HSS structure.

In the present work, we investigate the stability of the ground state against the single-site uniaxial anisotropy $D$. In analogy with the VBS state, if the ground state of the $S=1$ KHAF is the HSS state, it  should be also destroyed by easy plane single-site anisotropy, because the strong anisotropy projects out the $S^z=\pm1$ states in each site. We also investigate the combined effect of the  $\sqrt{3}\times\sqrt{3}$ lattice distortion and single site anisotropy $D$.

This paper is organized as follows. The model Hamiltonian is presented in the next section. The effect of single-site anisotropy is investigated in \S 3. The combined effect of  single-site anisotropy and lattice distortion is discussed in \S 4. The ground state phase diagram is also presented. The last section is devoted to summary and discussion.

\section{Model Hamiltonian}

We consider the anisotropic distorted $S=1$ KHAF given by,

\begin{eqnarray}
\label{ham2}
\H &=& \HA + \HB+ \HC+\sum_i D{S}^{z2}_{i},  \nonumber  \\
\H_{\alpha} &=& J_{\alpha}\sum_{<i,j> \in \alpha}\v{S}_{i} \v{S}_{j}, \nonumber
\end{eqnarray} 
\begin{figure}
\centerline{\includegraphics[width=7cm]{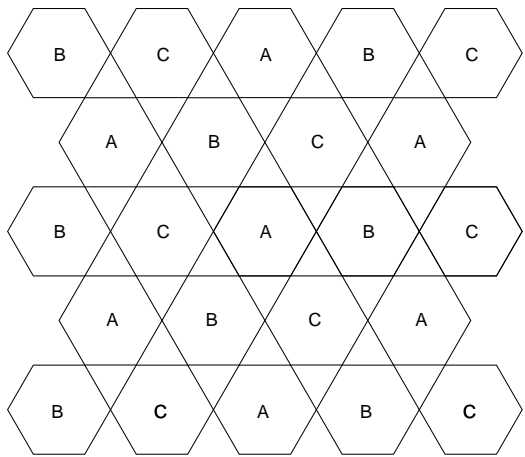}}
\caption{$\sqrt{3}\times\sqrt{3}$ distorted kagom\'e lattice. }
\label{fig2}
\end{figure}
where $\v{S}_{i}$ is the spin operator with $S=1$ and $\sum_{<i,j> \in \alpha}$  represent the summation over the bonds around the type-$\alpha$ $ (\alpha = \mbox{A, B} \ \mbox{or  C})$ hexgons, which are depicted in Fig. \ref{fig2}. We take $\JA=1$ and $\JB=\JC=\alpha$ with $0 <\alpha < 1$.

\section{Effect of Single-Site Anisotropy $D$}

Let us first concentrate on the case $\alpha=1$ and $D > 0$. The HSS state is obviously destroyed for $D >> J$, because  each spin can take only the single state $S_i^z=0$ and no HSS state can be constructed.  If the ground state of the uniform $S=1$ KHAF is the HSS state, a phase transition should therefore take place at an intermediate value of $D$. 
 
 Corresponding phenomenon is well established for  $S=1$ AFHC, in which the VBS state is destabilized by large easy plane single site anisotropy\cite{dnr,tasaki} resulting in the "large-$D$" phase. This observation elucidated that the ground state of the uniform $S=1$ AFHC has the VBS structure. Therefore, in the present case also, the presence of this transition is a strong support for the HSS picture of the uniform $S=1$ KHAF.

\begin{figure}
\centerline{\includegraphics[width=7cm]{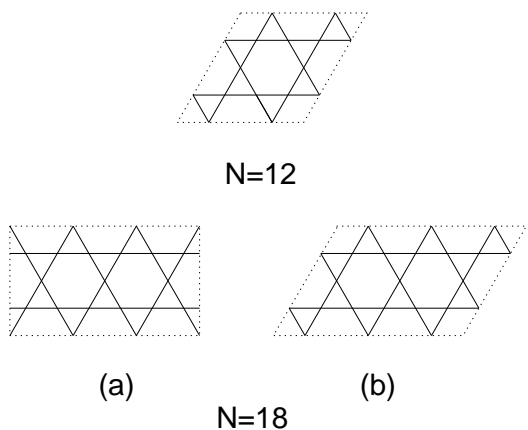}}
\caption{Clusters used for numerical diagonalization. }
\label{clusters}
\end{figure}

\begin{figure}
\centerline{\includegraphics[width=7cm]{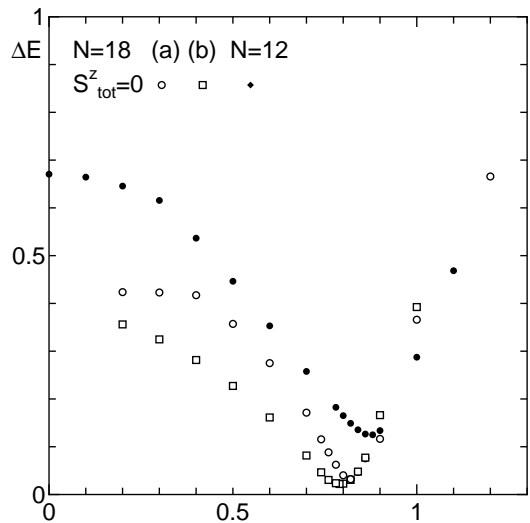}}
\caption{The $D$-dependence of the lowest energy gap with $S^z_{\rm tot}=0$ of $S=1$ anisotropic KHAF $N=18$ (cluster a), $N=18$ (cluster b) and $N=12$.}
\label{gapdis}
\end{figure}

To confirm the presence of the phase transition between the HSS and large-$D$ states, the numerical diagonalization calculation is carried out for the finite size clusters with $N=12$ and 18. The clusters used for the calculation are shown in Fig. \ref{clusters}. The gap  $\Delta E$ between the ground state and the first excited state with $z$ component of the total spin $S^z_{\rm tot}=0$ is plotted in Fig. \ref{gapdis} against $D$ for $N=12, 15$ and 18. The excitations with higher spins have larger excitation energy. It is evident that the gap has a minimum at $D \sim 0.82$ for $N=18$ (a) type cluster, at $D \sim 0.79$ for $N=18$ (b) type cluster and at $D \sim 0.88$ for $N=12$ cluster. 
Although we cannot analyze the system size dependence of the gap in detail due to the limitation of the size, this result indicate the presence of the phase transition in the thermodynamic limit around $D \simeq 0.8$.

\section{Ground State Phase Diagram}

\begin{figure}
\centerline{\includegraphics[width=7cm]{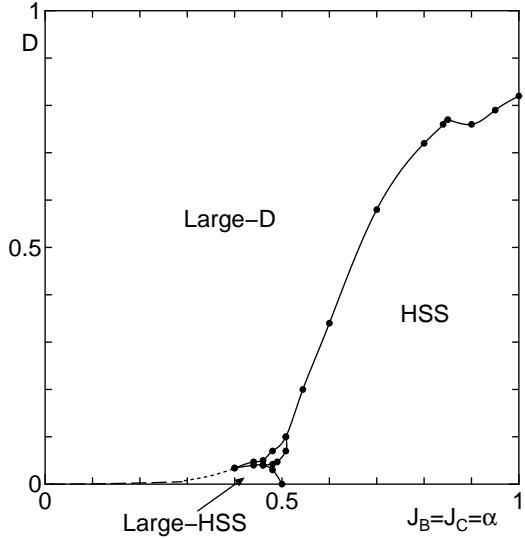}}
\vspace{5mm}
\caption{The phase diagram expected from the numerical data with  $N=18$. The solid and dotted lines are guides for eye. The broken line is the result in the strong distortion limit $\alpha << 1$.}
\label{phase}
\end{figure}

Let us consider the general case of $D > 0$ and $0 < \alpha <1$. In this case, the unit cell contains 9 spins and only $N=18$ cluster of type (a) is compatible with this type of lattice distortion with periodic boundary condition. Therefore we cannot check the size dependence of the energy gaps. We nevertheless roughly estimate  the phase boundary from the point at which the energy gap takes a minimum for this cluster.  The qualitative phase diagram expected from these finite size data is shown in Fig. \ref{phase}. Strictly speaking, we find a slight reentrant behavior around $\alpha \sim D \sim 0.8$ and a small intermediate phase between the HSS and large-$D$ phases around $\alpha \sim 0.5$ and $D \sim 0.1$. Considering the crudeness of the method used for the determination of the phase boundary, it is not clear whether these singular parts of the phase boundary are physically meaningful or artifacts of the finite size calculation.

\begin{figure}
\centerline{\includegraphics[width=7cm]{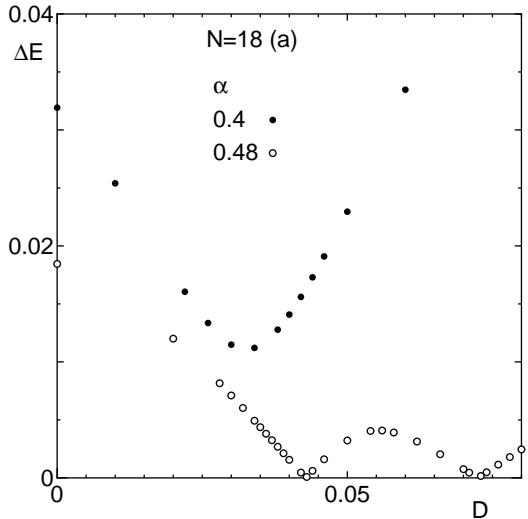}}
\vspace{5mm}
\caption{The $D$-dependence of the lowest energy gap with $S^z_{\rm tot}=0$ of $S=1$ distoorted anisotropic KHAF with $N=18$ (cluster a) at $\alpha=0.4$ and 0.48. The ground state is the large-HSS state for $D=0$.}
\label{smd}
\end{figure}

Even for  $\alpha \simleq 0.5$, where the ground state for $D=0$ is the large HSS state, the gap takes a minimum with the increase of $D$ as shown in Fig \ref{smd}. Although the $N=18$ cluster does not support the large-HSS structure, the presence of the minimum of the gap at finite $D$ implies that the "alive" spins are killed by $D$ term at this value of $D$. Therefore we may roughly estimate the large-HSS-large-$D$ transition point from this minimum point of the energy gap.  The double minima in the case of $\alpha=0.48$ implies the possible presence of the intermediate phase. 

It should be remarked that the large-HSS phase is very fragile against $D$. As seen from the physical picture of the large-HSS phase explained in ref. \citen{kh3}, the effective interactions between the "alive" spins are of the order of $\alpha^2$ for small $\alpha$. Therefore we may expect that the large-HSS state is destabilized around $D \sim O(\alpha^2)$. As explained in ref. \citen{kh3}, the dominant effective interaction between the 'alive' spins is the next nearest neighbour interaction of the enlarged kagom\'e lattice of the 'alive' spins. If we retain only this term and $D$-term, the large-$D$-large HSS transition point can be estimated from that of the uniform $S=1$ KHAF for small $\alpha$. The result is $D_c \simeq 0.06\alpha^2$. This is plotted by the broken line in Fig. \ref{phase}. The solid and dotted  curves are the guide for eye.

\section{Summary and Discussion}

\begin{center}
\begin{table}
\caption{Correponding phases of the $S=1$ KHAF and $S=1$ antiferromagnetic Heisenberg chain (AFHC).}
\label{table1}
\begin{tabular}{@{\hspace{\tabcolsep}\extracolsep{\fill}}ccc} 
 \hline
 & $S=1$ AFHC & $S=1$ KHAF \\ \hline\hline
small $D$ small $\delta (= 1-\alpha)$ & VBS-phase & HSS-phase \\ \hline
large $\delta$ & dimer-phase & large-HSS phase \\ \hline
large $D$ & large-$D$ phase & large-$D$ phase \\ \hline
\end{tabular}
\end{table}
\end{center}

In summary, we have found that the ground state of the distorted $S=1$ KHAF with single site anisotropy $D$ undergoes the phase transition to the large-$D$ phase for large enough $D$. The presence of this phase transition is consistent with the HSS picture for the isotropic $S=1$ KHAF which is analogous to the VBS picture of the Haldane phase\cite{kh1,aklt}. In this context, this transition corresponds to the Haldane-large-$D$ phase transition in the $S=1$ AFHC\cite{dnr,tasaki}. 

The effect of the $\sqrt{3}\times\sqrt{3}$ lattice distortion on this phase transition is also investigated. The obtained phase diagram is remniscent of the corresponding phase diagram of the $S=1$ AFHC with dimerization and single-site anisotropy\cite{chen,tone}. The correspondence between these two models is summarized in table \ref{table1}. The important difference is, however, that there exists a phase transition between the large HSS phase and the large-$D$ phase unlike the case of  $S=1$ AFHC in which no phase transition between the dimer phase and large-$D$ phase takes place\cite{chen,tone}.

The numerical calculation is performed using the HITAC SR8000 at the Supercomputer Center, Institute for Solid State Physics, the University of Tokyo and the HITAC SR8000 at the Information Processing Center, Saitama University.  The numerical diagonalization program is based on the TITPACK ver.2 coded by H. Nishimori and KOBEPACK/1 coded by T. Tonegawa, M. Kaburagi and T. Nishino. This work is supported by the Grant-in-Aid for Scientific Research from the Ministry of Education, Culture, Sports, Science and Technology, Japan.


\begin{thebibliography}{999}
\bibitem{rev} For a review see {\it Magnetic Systems with Competing Interactions: frustrated spin systems}, edited by H. T. Diep (World Scientific, Singapore, 1994) and references therein.
\bibitem{ze1} C. Zeng and V. Elser: Phys. Rev. {\bf B42}  8436 (1990).
\bibitem{ey1} N. Elstner and A. P. Young: Phys. Rev. {\bf B50}  6871 (1994).
\bibitem{nm1} T. Nakamura and S. Miyashita:  Phys. Rev. B52 9174 (1995).
\bibitem{wal1} Ch. Waldtmann, H.-U. Everts, B. Bernu, C. Lhuillier, P. Sindzingre, P. Lecheminant and L. Pierre: Eur. Phys. J. {\bf B2} 501 (1998). 
\bibitem{mila1} F. Mila:  Phys. Rev. Lett. {\bf 81}, 2356 (1998)
\bibitem{sin1} P. Sindzingre, G. Misguich, C. Lhuillier, B. Bernu, L. Pierre, Ch. Waldtmann and H.-U. Everts:  Phys. Rev. Lett. {\bf 84} 2953 (2000).
\bibitem{kh1} K. Hida: J. Phys.Soc. Jpn. {\bf 69} 4003 (2000).
\bibitem{kh2} K. Hida: J. Phys.Soc. Jpn. {\bf70} 3673  (2001).
\bibitem{kh3} K. Hida: J. Phys.Soc. Jpn. {\bf71} 1027  (2002).
\bibitem{aklt} I. Affleck, T. Kennedy, E. H. Lieb and H. Tasaki: Commn. Math. Phys. {\bf 115} (1988) 477; Phys. Rev. Lett. {\bf 59} 799 (1987).
\bibitem{yama1} Y. Yamashita and K. Ueda: Phys. Rev. Lett. {\bf 85} 4960 (2000).
\bibitem{wada1} N. Wada, T. Kobayashi, H. Yano, T. Okuno, A. Yamaguchi and K. Awaga: J. Phys. Soc. Jpn. {\bf 66}  961 (1997).
\bibitem{awaga1} K. Awaga, T. Okuno, A. Yamaguchi, M. Hasegawa, T. Inabe, Y. Maruyama and N. Wada: Phys. Rev. {\bf B49} 3975 (1994) .
\bibitem{wata1} I. Watanabe, N. Wada, H. Yano, T. Okuno, K. Awaga, S. Ohira, K. Nishiyama, K. Nagamine: Phys. Rev. {\bf B58} 2438 (1998).
\bibitem{dnr} M. den Nijs and K. Rommelse: Phys. Rev. B 40, 4709 (1989). 
\bibitem{tasaki} H. Tasaki: Phys. Rev. Lett. 66, 798 (1991). 
\bibitem {chen} W. Chen, K. Hida and B. C. Sanctuary:  J. Phys. Soc. Jpn. {\bf 69} (2000) 237. 
\bibitem {tone} T. Tonegawa, T. Nakao and M. Kaburagi: J. Phys. Soc. Jpn. {\bf 65} (1996) 3317. 
\end{thebibliography}
\end{document}